\documentclass{elsart}
\usepackage[square,sort,comma,numbers]{natbib}
\usepackage{graphicx}
\usepackage{amssymb}

\begin{document}

\begin{frontmatter}

\title{Background identification and suppression for the measurement of (n,$\gamma$) 
reactions with the DANCE array at LANSCE}

\author[lanl]{R.~Reifarth\corauthref{cor}},
\ead{reifarth@lanl.gov}
\author[lanl]{T.A.~Bredeweg},
\author[mines]{A.~Alpizar-Vicente},
\author[lanl]{J.C.~Browne},
\author[lanl]{E.-I.~Esch},
\author[mines]{U.~Greife},
\author[lanl]{R.C.~Haight},
\author[mines]{R.~Hatarik},
\author[lanl]{A.~Kronenberg},
\author[lanl]{J.M.~O'Donnell},
\author[lanl]{R.S.~Rundberg},  
\author[lanl]{J.L.~Ullmann}, 
\author[lanl]{D.J.~Vieira}, 
\author[lanl]{J.B.~Wilhelmy},
\author[lanl]{J.M.~Wouters}

\address[lanl]{Los Alamos National Laboratory, Los Alamos, New Mexico, 87545, USA}
\address[mines]{Colorado School of Mines, Golden, Colorado, 80401, USA}
\corauth[cor]{Corresponding author:}

\begin{abstract}
In the commissioning phase of the DANCE project (Detector for Advanced Neutron Capture 
Experiments) measurements  have been performed with special emphasis on the identification 
and suppression of possible backgrounds for the planned (n,$\gamma$) experiments. 
This report 
describes several background sources, observed in the experiment or anticipated from 
simulations, which will need to be suppressed in this and in similar 
detectors that are planned at other facilities. First successes are documented in 
the suppression of background from scattered neutrons captured in the detector as well as
from the internal radiation. 
Experimental results and simulations using the GEANT code are compared.

\end{abstract}


\end{frontmatter}

\section{Introduction}
\label{}
DANCE (Detector for Advanced Neutron Capture Experiments) is a 4$\pi$ detector array that 
consists of up to 160 elements of barium fluoride crystals. It is designed to study 
capture reactions on small quantities of radioactive isotopes, which are of interest 
to studies in nuclear astrophysics and stockpile stewardship science. DANCE is 
located on the 20~m neutron flight path 14 (FP14) at the Manuel Lujan Jr. Neutron Scattering 
Center at the Los Alamos Neutron Science Center (LANSCE) \cite{LBR90}. The neutrons are produced via 
spallation 
reactions caused by an 800 MeV proton beam hitting a tungsten target with typical beam
currents of 100~$\mu$A. Depending on the flight path, the fast neutrons are moderated by water 
or other moderators. FP14 is 
designed to view only the uncoupled water moderator and not the tungsten target itself. 
A direct neutron capture measurement in an environment like this faces 
three major types of background:

1. Time-independent background, which is not correlated with the neutron beam. Natural radioactivity or, 
in case of BaF$_2$ detectors, intrinsic radioactivity are prominent examples. If the sample is radioactive, 
it will introduce another source of time-independent background.\\
2.	Time-dependent background, which is every background component that is correlated with the time structure 
of the beam, but does not scale with the size of the sample in the beam. Examples are neutrons scattered 
from other experiments, or any particles originating at the neutron production area. \\
3.	Sample-related background due to two processes:\\
3a)	(n,n) reactions, where the neutrons are eventually captured in the surrounding material 
(in case of DANCE mainly BaF$_2$) and create a similar signature as capture in the sample. \\
3b)	($\gamma$,X), where X includes all the possible interaction mechanisms between matter and photons 
(photo-effect, Compton scattering, pair production etc.). Most of the interactions are with the 
electrons in the sample material.

While (1) and (2) are usually determined by simple beam on/off and sample in/out experiments, 
more sophisticated measurements are needed to determine the sample related background (3).  In order to 
determine the background due to scattered neutrons (3a), either the time structure or the energy 
information of the total $\gamma$-ray energy released is used. Classical (n,$\gamma$) experiments were 
optimized to have detectors with very low neutron sensitivity (e.g. C$_6$D$_6$) and could only measure the 
resonant 
part of the cross section. All the events, which appeared at a time of flight between resonances 
were considered to be background. The neutron scatter background was not measured, because it
was assumed to be negligible. However, as it turned out, this method works fine only if the capture-to-scatter 
cross section ratio is not too small and the required uncertainties are on the order 
of 10\%. For higher accuracies or less favorable cases, a better method is to detect 
all the $\gamma$-rays emitted during an event \cite{WGK90a}. 
This means using detector material with higher gamma as well as neutron sensitivity. The number of 
events due to captures of scattered neutrons in the detection system will therefore be increased. But 
this method has the advantage that neutron captures on different isotopes, in particular on the sample and
detector/structure materials, can be discriminated on the basis of the total energy released, or the so-called 
Q-value. 
Further background reduction might be necessary
because of an unfavorable signal to background ratio. Applying appropriate cuts on total energy and 
multiplicity 
can therefore lead to a much improved signal-to-background ratio.  
Additionally, the significantly higher detection efficiency of such setups allows the use of 
smaller samples. 

Background due to $\gamma$-rays (3b), which are produced at the neutron production site or from
neutron capture on beam line components, can arrive at the sample position at the same time of 
flight as the neutrons under investigation. Without a sample in the beam, these gammas 
would not be detected, but since they interact with the sample (mainly via Compton-scattering 
and pair production), secondary gammas will be detected. Obviously this background scales with 
the sample size. Most of the $\gamma$-rays produced have energies below 3~MeV and the energy deposition in a
calorimetric detection system will therefore be much smaller than for the reactions under investigation. 

Additionally, events due to scattered high-energy neutrons at earlier times might appear during the  
TOF-window under investigation, since most of the neutrons will be moderated before the final capture. 
Disentangling between (3a) and (3b) is therefore a challenge for modern high accuracy (n,$\gamma$) 
experiments. 
This report describes the attempts of the DANCE collaboration to identify and suppress different 
background components during the different commissioning phases.

\section{Experimental Setup}
\label{}

DANCE is designed as a high efficiency, highly segmented 4$\pi$ BaF$_2$ detector for calorimetrically 
detecting 
$\gamma$-rays following a neutron capture. The initial design work is described in \cite{HRF01}. 
For practical reasons 
the detector modules do not really cover the entire solid angle. The design of the detector
is such that a full 4$\pi$ array would consist of 162 crystals of four different shapes, 
each shape covering the same solid angle \cite{HSD79}. Two of the 162 
crystals are left out in order to leave space for the neutron beam pipe. 
Depending on the experiment, 
one crystal can be replaced by a sample changer mechanism, which makes it possible to exchange up to 
3~samples without closing the beam shutter and breaking the vacuum of the beam pipe. Thus the 
full array is designed to host 159 or 160 out of 162 possible BaF$_2$ crystals. The dimensions of
the bare crystals are
designed to form a BaF$_2$ shell with an inner radius of 17~cm and a thickness of 15~cm.

\begin{figure}
\begin{center}
\includegraphics[width=6cm]{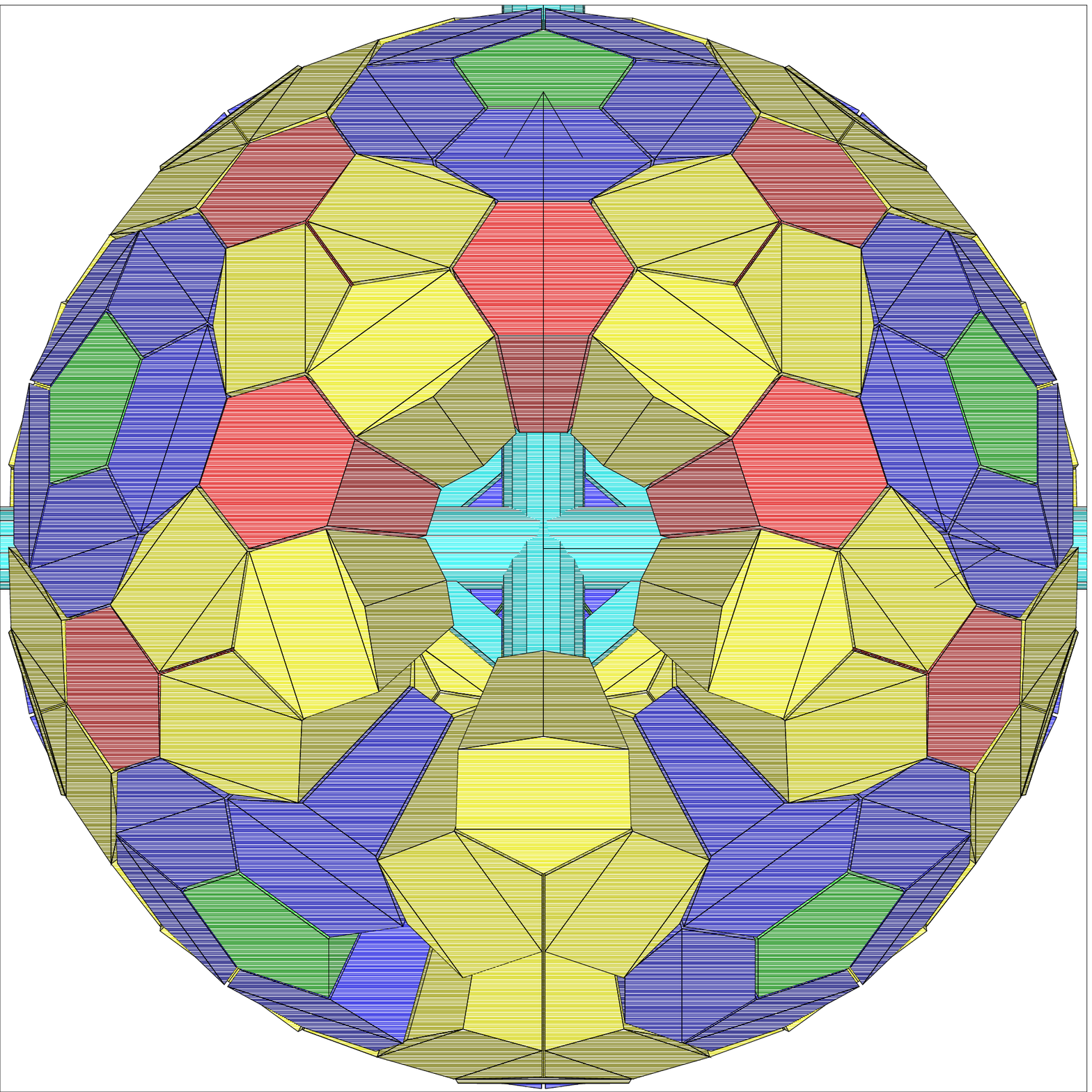}
\includegraphics[width=6cm]{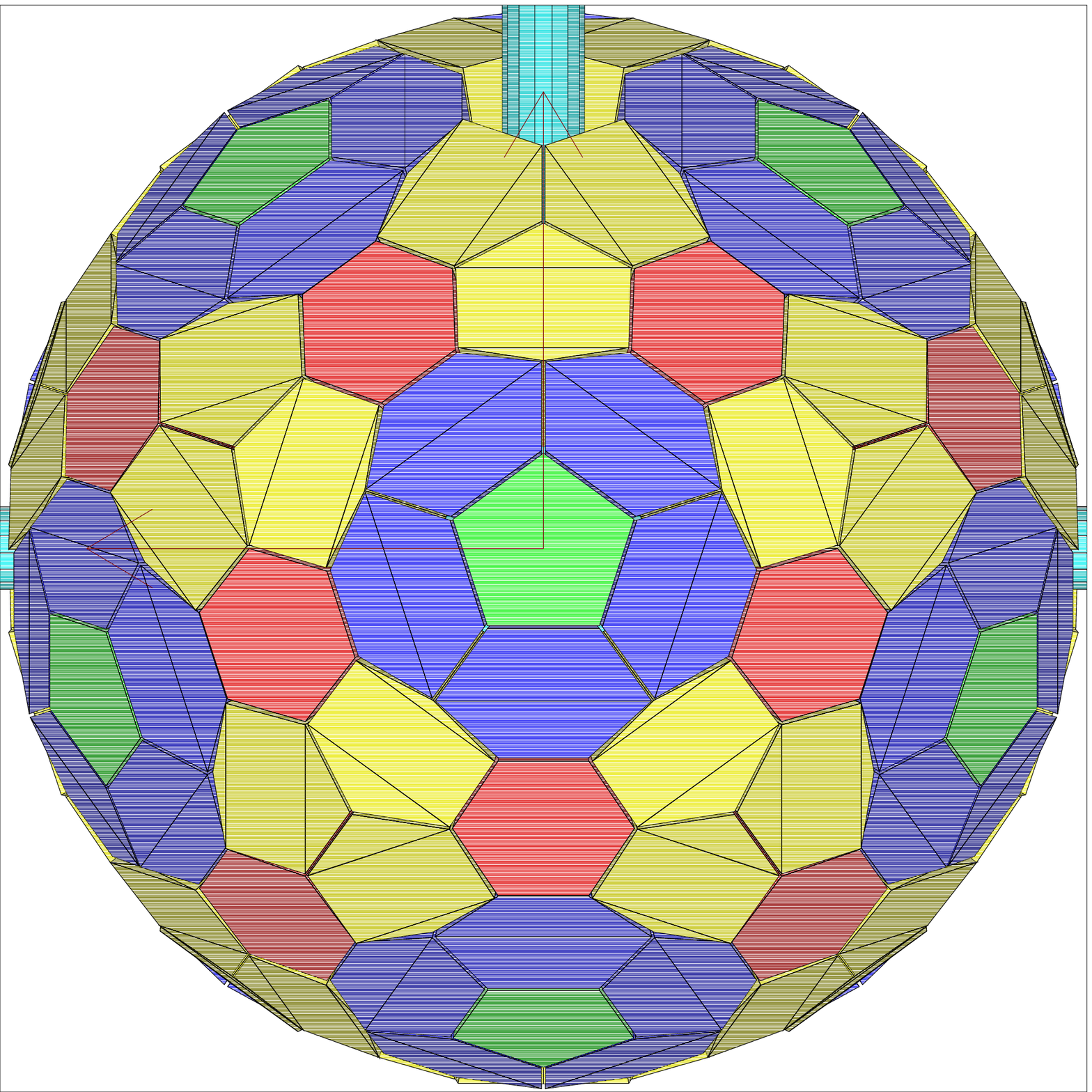}
\includegraphics[width=6cm]{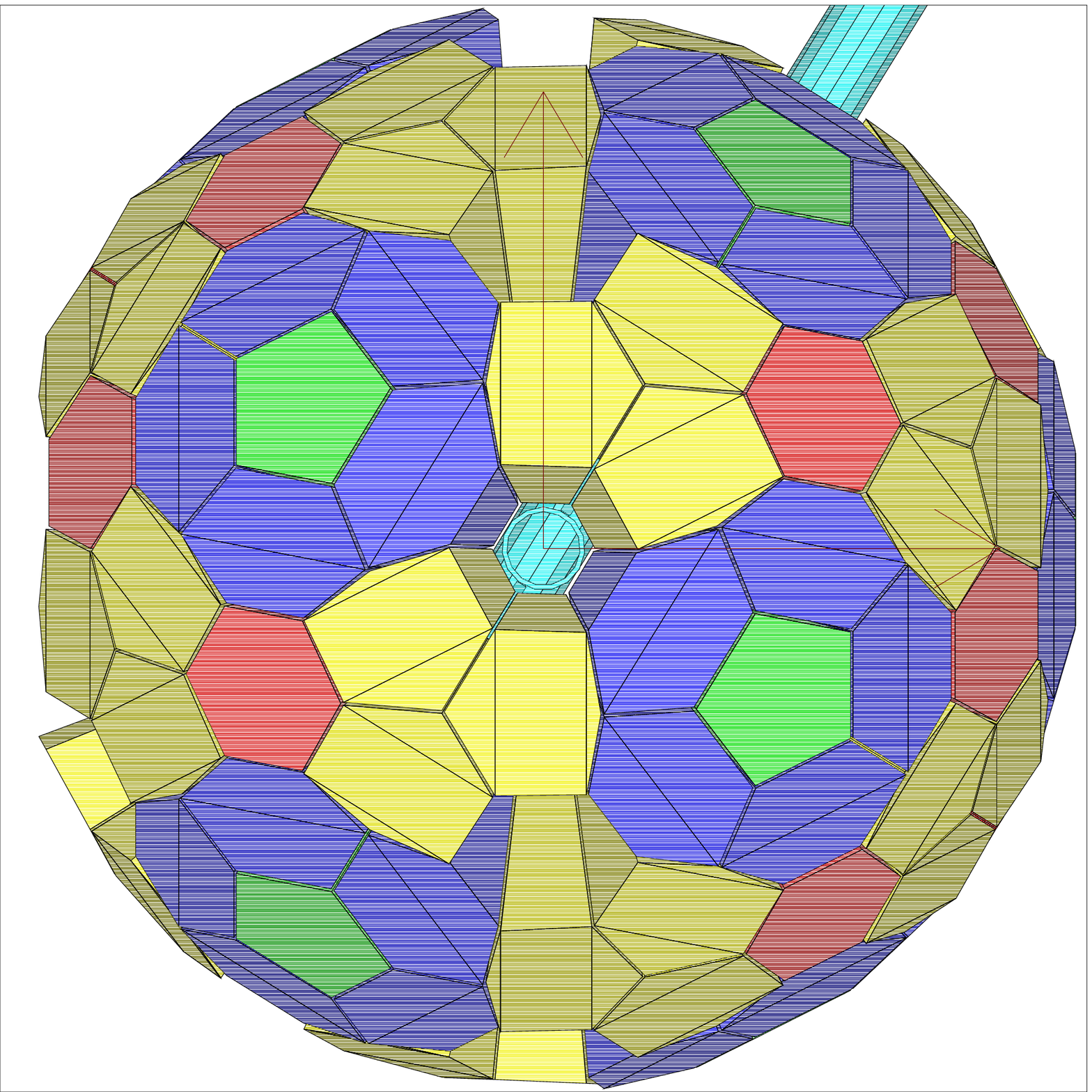}
\includegraphics[width=6cm]{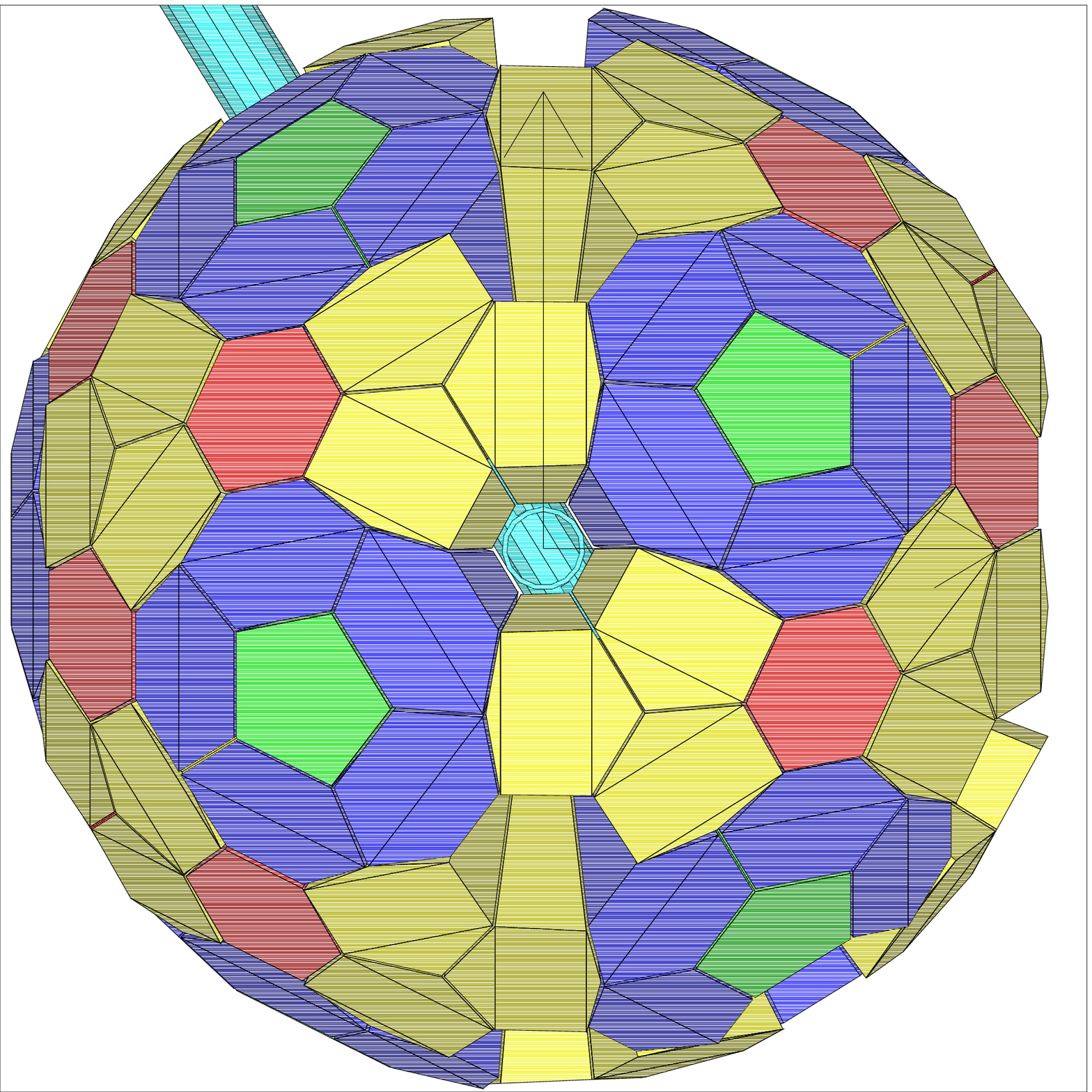}
\end{center}
\caption{Views of the simulated setup: from left / right / downstream / upstream 
(relative to beam direction; clockwise starting from top left). Each color corresponds to a
different crystal type (A - green, B - dark blue, C - yellow, D - red). The missing crystals allowing the beam 
pipe (light blue) to be seen were either not connected or not in place during the runs carried out during the
commissioning phase 2002/2003.}
\label{fig_sim}
\end{figure}

During the first commissioning phase of this project, from November 2002 until January 2003, only 147 
crystals were installed in the array and only 141 were actually connected. 
The main part of this report concentrates on the
results obtained with 141 active crystals. Only Sect.~\ref{gamma_induced} contains data taken with the full
array of 159/160~crystals. 
In order to guide the commissioning process, intensive simulations using the detector simulation
package GEANT~3.21 were performed \cite{GEA93}. Fig.~\ref{fig_sim} illustrates the conditions as they were simulated. 
The simulations were designed to be as close to the experimental conditions as possible. 
Therefore, the 6 mounted crystals that 
were not connected were included in the simulations as passive detectors, but were not drawn in the 
figure in order to increase the visibility of the interior of the ball. Four supporting 
structure pieces were also included. These pieces are welded onto the cross of the beam line inside the 
ball in order to increase the mechanical stability (see top left part of Fig.~\ref{fig_sim}). 
The DANCE array is divided into two halves, 
left-hand side and right-hand side relative to the neutron flight direction. The position of almost 
all of the crystals, which were not mounted at that time, was on the equatorial ring joining the two halves.
The response of the array to neutrons and gamma rays was studied using the Monte Carlo code GEANT 3.21 
including the GCALOR package for low energy neutron transportation \cite{GEA93,ZeG94}. 
The inner radius of a closed sphere with bare crystals would be 17~cm and the crystals have a length of 
15~cm. The crystals are supported by a spherical structure made of aluminum with an inner radius of 
49.7~cm and an outer radius of 53.5~cm (1.5~inches thickness). Each crystal is wrapped in a PVC 
foil of 0.7~mm thickness and glued to a photo multiplier tube (PMT). In reality as well as in the simulations, 
the crystals had to be moved outwards by 1~cm in order to leave space for this wrapping. Together with the
fact that only 160~crystals can be used, the actual solid angle covered by the detector is reduced to about 
3.6$\pi$. The crystal-PMT unit was put into an aluminum housing in a way that there is no aluminum 
between the crystals or between crystals and sample, while the PMT is
surrounded by an aluminum-housing, which was  
finally mounted to the surrounding supporting structure. The supporting structure as well as the aluminum 
housings were included in the simulations. Due to restrictions in the number of defined volumes 
only a simplified PMT could be included. The simplification was that the material of the PMT 
(mainly Co and Ni contained in the magnetic shielding) was mixed into the material of the Al-housings. 
In order to make the simulations as realistic as possible, the beam pipe including the cross for the 
sample changer were included in the simulations. 

Previous GEANT simulations \cite{HRF01} have predicted that the background due to neutrons, which are 
scattered at 
the sample and eventually captured in the surrounding material, can be significantly reduced by a 
spherical $^6$LiH shell between the sample and the BaF$_2$ detectors. Some measurements carried 
out during the commissioning
phase had such a $^6$LiH moderator with an inner radius of 10.5 cm and an outer radius of 16.5~cm
surrounding the target location. 
Therefore, most of the simulations described here were done both with and without such a moderator.
The simulated $^6$LiH absorber had a density of 0.85 g/cm$^3$ and a isotopic composition of 100\% $^6$Li.

In order to check the effect of scattered neutrons, 
a set of simulations including the walls of the detector cave (made of borated polyethylene (5\%)) and
aluminum windows up- and downstream of the DANCE array has been carried out 
(see Section~\ref{time-dependent}).
The threshold behavior, the energy resolution and the air gap between the crystals were included
in the simulations in an attempt to fit the measured spectra for various calibration source spectra. 

\section{Background components}
\label{}

\subsection{Count rate estimate}
\label{count_estimate}

In this section a rough estimate of the expected count rate during a typical experiment 
will be presented. The neutron production target in the Lujan Center is typically irradiated 
by a pulsed 800~MeV proton beam with an average current of 100~$\mu$A and a repetition rate of 20~Hz.
Unless otherwise stated this section and all of the figures in the following sections assume 
a proton pulse of these specifications. Under these conditions the neutron
flux at the sample position inside DANCE follows approximately a 1/$E$ law in the neutron
energy region of interest 1~eV~$<~E~<$~1~MeV. This results in a constant number 
of about 3$\cdot$10$^5$~s$^{-1}$ neutrons per energy decade, or 1.5$\cdot$10$^4$ 
neutrons per energy decade per proton pulse. 

Assuming a 1/$\sqrt{E}$ energy dependence for the neutron capture cross
section, the neutron capture rate in the sample is expected to be constant over the duration 
of the neutron pulse. If the (n,$\gamma$) cross section is 
$\sigma(E)$~=~$c/\sqrt{E}$ and $N_{sample}$ is the number of sample atoms, 
the instantaneous neutron capture rate at any time during the duration of the neutron
pulse would be:
\begin{eqnarray*}
\frac{\mbox{d}C}{\mbox{d}t} & = & \frac{\mbox{d}\Phi}{\mbox{d}E}\cdot\frac{\mbox{d}E}{\mbox{d}t}
								  \cdot\sigma(E){\cdot}N_{sample} \\
							& = & 2.7\cdot10^8{\cdot}c{\cdot}N_{sample}~\frac{1}
								{\sqrt{\mbox{keV}}~\mbox{s cm}^2}~.
\end{eqnarray*}
Inserting typical values of $c~=~1~\mbox{barn} \sqrt{\mbox{keV}}$ and $N_{sample}~=~6\cdot10^{18}$ 
(1~mg with atomic mass 100) gives 1.65$\cdot10^3$ neutron capture per second, which will be detected with 
nearly 100\% efficiency. Most of
the neutron capture events will cause a signal with summed $\gamma$-energy of $\pm$500~keV 
around the Q-value of the 
reaction. Based on these assumptions an instantaneous capture rate of $10^3$~s$^{-1}$ distributed over 
a $\gamma$-energy range of 1~MeV will be used for comparison with the achieved background 
levels in later sections.

\subsection{Time-independent background}
\label{}

Time-independent background is not correlated with the time structure of the neutron beam. 
In our case natural radioactivity, the intrinsic radioactivity of the BaF$_2$ detectors 
and that of a radioactive sample have to be considered. 

Natural $\gamma$-activity, like $^{40}$K and $^{208}$Tl, with $\gamma$-ray
energies of 1465~keV and 2615~keV, respectively, can be discriminated from neutron capture events
using information about the gamma energy deposited in the detector. 
The neutron binding energy is above 5~MeV for most of the isotopes under investigation, which 
is well separated from the typical $\gamma$-ray energies of natural decays.
This means that most of the events due to neutron capture on the sample appear at much higher channels 
in the sum energy spectrum. As it will become clearer in later sections, it is still desirable to understand
and - ideally - reduce or discriminate those lower-energy background components by other means than the 
energy information. The main reason for that is the fact that DANCE does not have a 100\% detection efficiency
for $\gamma$-rays and consequently a significant part of the (n,$\gamma$) events deposit therefore less energy 
than the
Q-value. For high accuracy measurements, this low energy tail of the energy distribution needs  
to be well understood and, if possible, measured with low background.  

\begin{figure}
\begin{center}
\includegraphics[width=.9\textwidth]{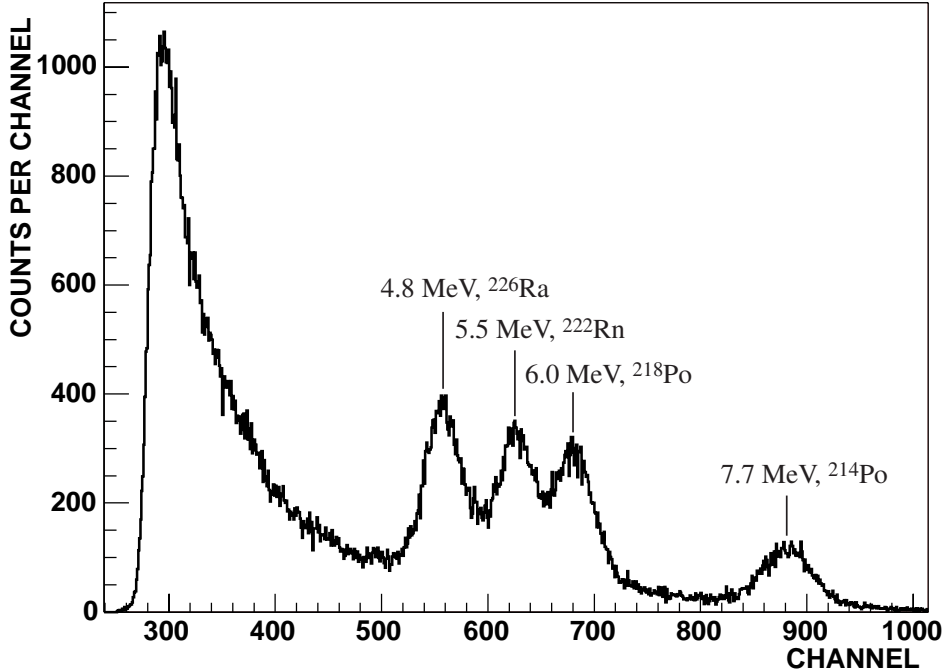}
\end{center}
\caption{The picture shows an $\alpha$-background spectrum for a single crystal without beam. The labeled peaks
correspond to different $\alpha$'s from the $^{226}$Ra decay chain.}
\label{fig_a_bkgd}
\end{figure}

\begin{figure}
\begin{center}
\includegraphics[width=.9\textwidth]{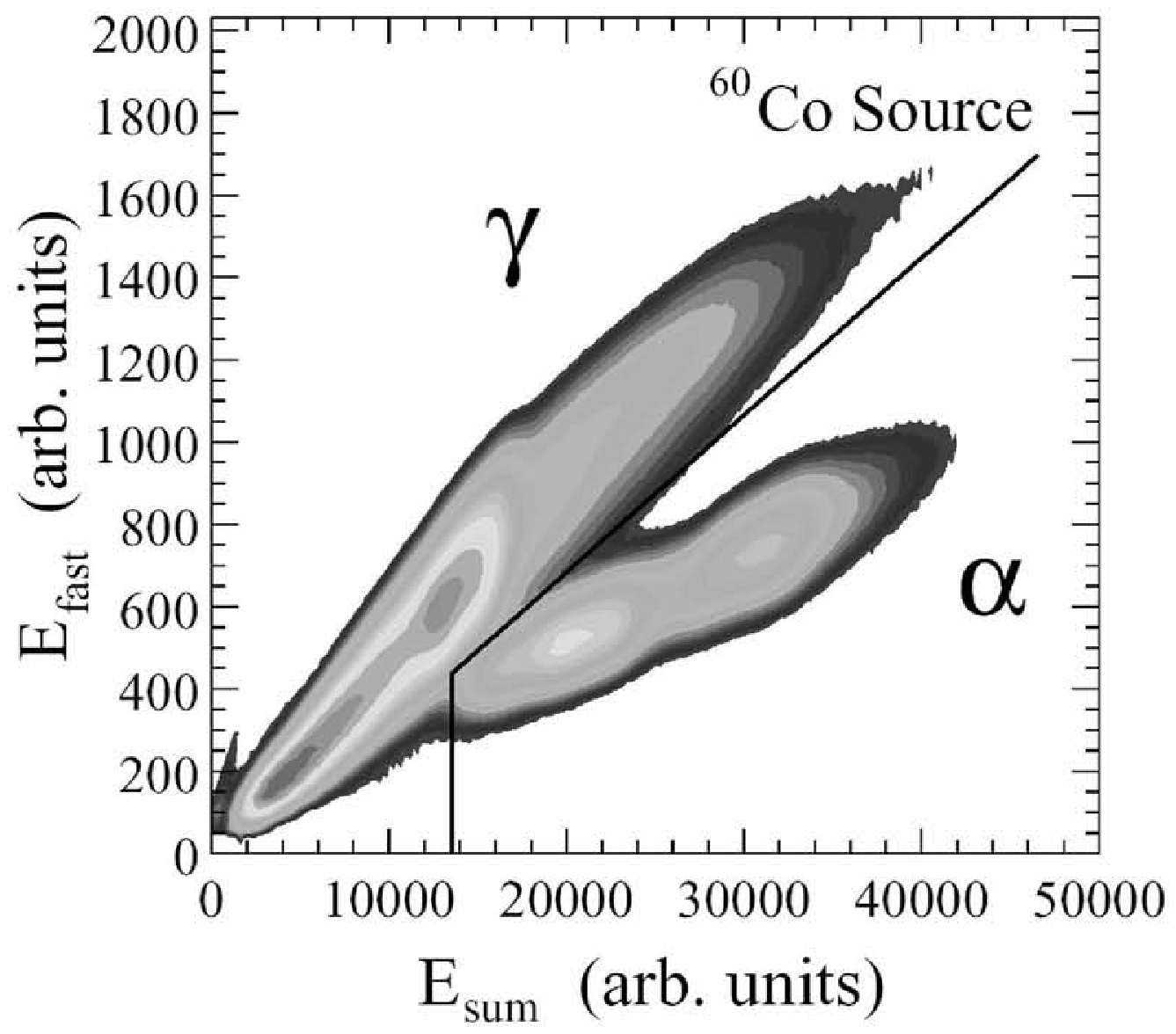}
\end{center}
\caption{A two-dimensional plot of the fast versus the sum of fast and slow 
component of the BaF$_2$ signal for a $^{60}$Co calibration source
is shown.
The lower right part is cut out in order to discriminate the intrinsic $\alpha$-background
and gain an improved signal to noise ratio.}
\label{fig_pid}
\end{figure}

The intrinsic radioactivity of the BaF$_2$ crystals originates from the $\alpha$-decay chain of the chemical
homologue $^{226}$Ra, see Fig.~\ref{fig_a_bkgd}. Typical activities are 0.2~Bq/cm$^3$ material \cite{WGK90a}. 
Most of the crystals used for DANCE show an intrinsic $\alpha$-activity between 150 and 250~Hz. The 
scintillation light of BaF$_2$ crystals has a short ($\approx$~0.6~ns) and a long ($\approx$~0.6~$\mu$s)  
component. The intensity ratio between the two components is radiation type dependent, which allows for 
particle
identification if both components are measured. While the emitted $\alpha$-particles deposit their energy 
almost
exclusively through the slow component, about 10\% of the energy of the $\gamma$-rays feeds the fast 
component. 
Therefore a comparison of the fast and slow component integrals provides a powerful
tool to reject the background due to $\alpha$-activity. In our experiment each BaF$_2$ signal is recorded with
waveform digitizers (Acqiris Corp., model DC-265) and integrated with two different integration times, 
each matching either the short
or the long decay time. Fig.~\ref{fig_pid} shows the results of this method applied to a run carried out with
a $^{60}$Co calibration source in the center of the DANCE array. 
Applying appropriate cuts results in an almost complete suppression of the $\alpha$-background. 

\begin{figure}
\begin{center}
\includegraphics[width=.9\textwidth]{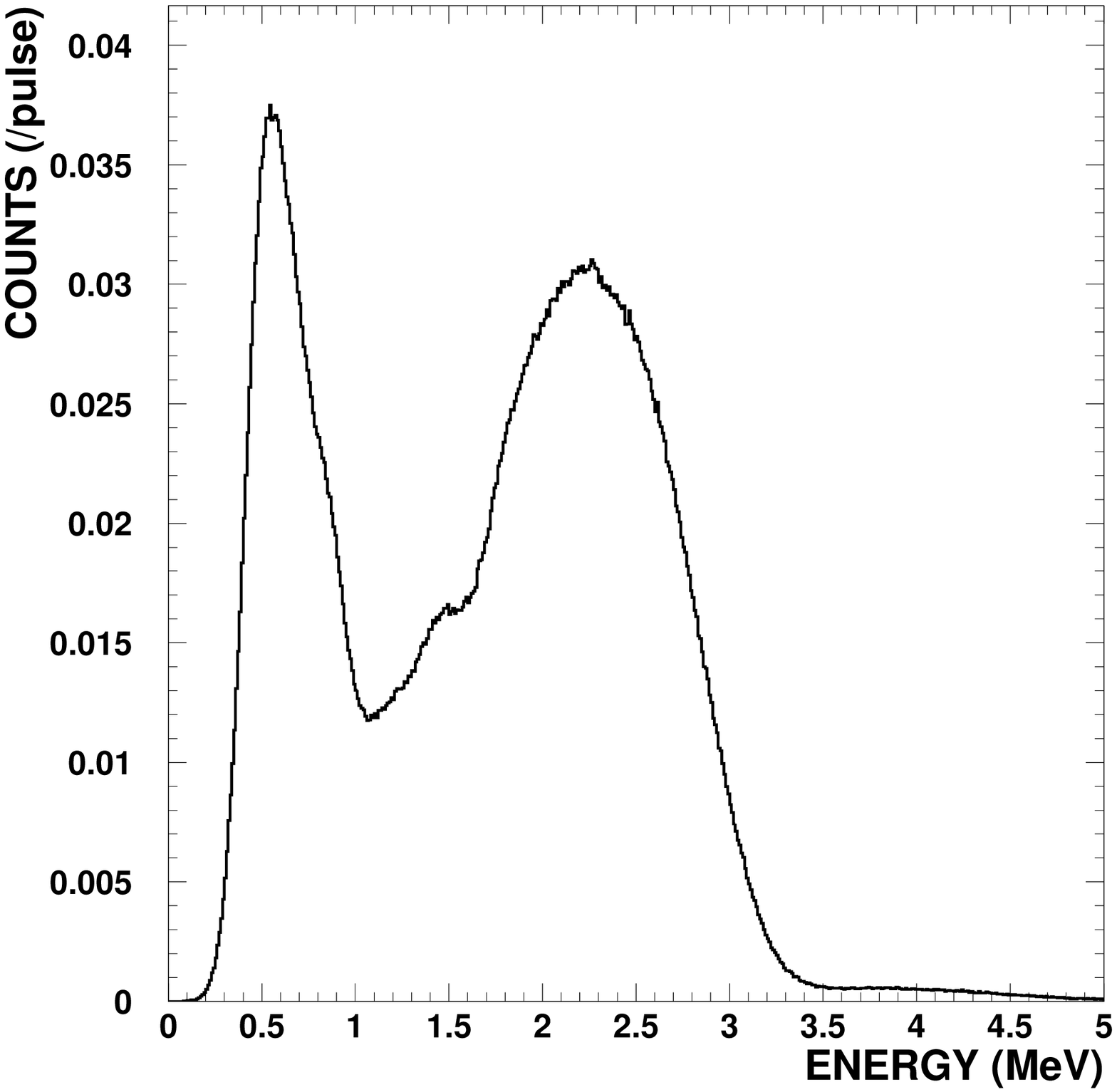}
\end{center}
\caption{Measured background for the DANCE array (160~crystals) without beam. Shown is the total energy for 
all events with detector multiplicity 2 or higher. The cut on multiplicity effectively eliminates 
events arising from internal $\alpha$-decays (e.g. Fig.~\ref{fig_a_bkgd}). Histogram binning is 
10~keV/channel.
One pulse corresponds to 2~ms looking time. The rate estimate given in Sect. \ref{count_estimate} would be
0.02 in this plot. }
\label{fig_bg}
\end{figure}

\begin{figure}
\begin{center}
\includegraphics[width=.9\textwidth]{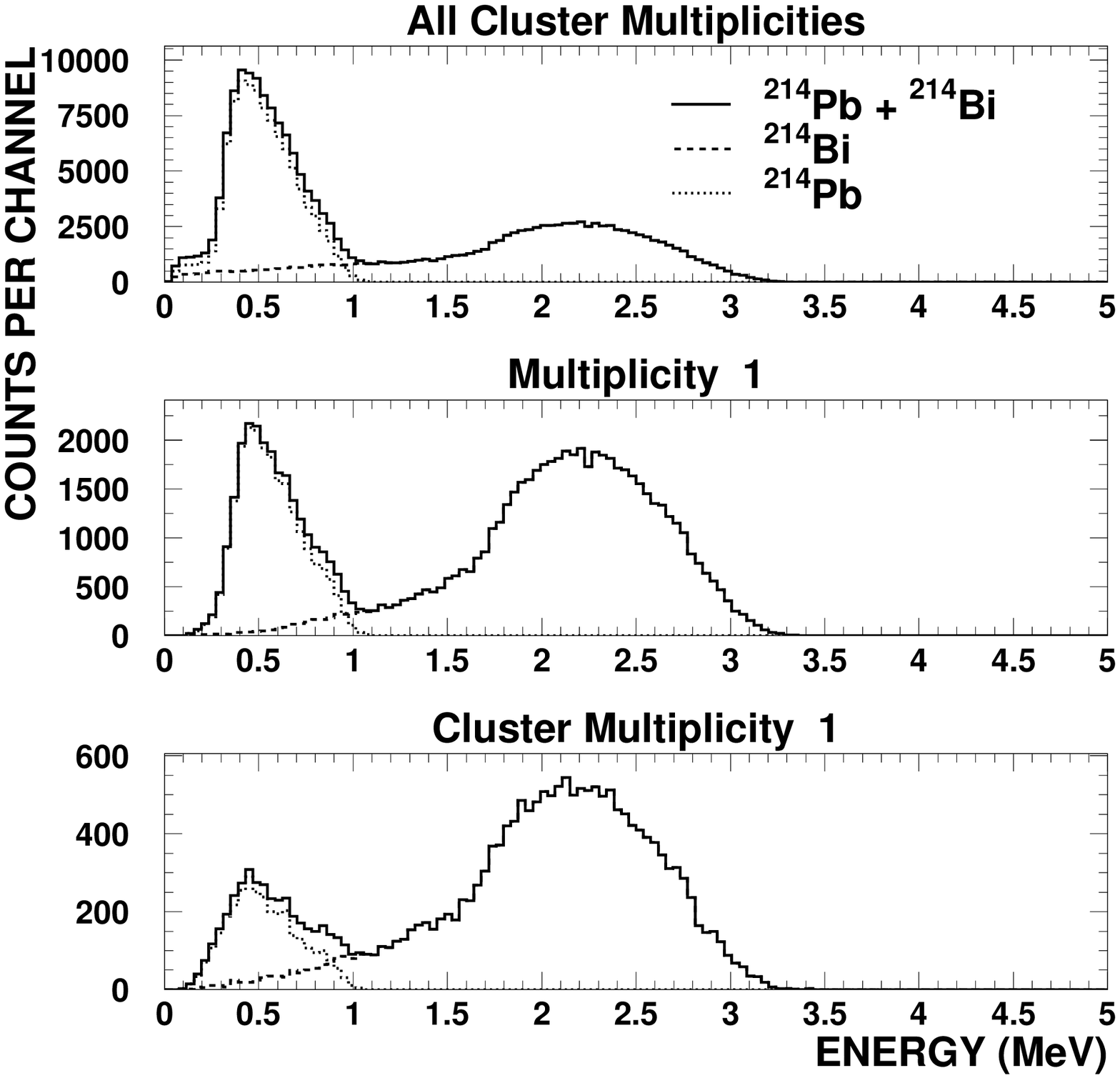}
\end{center}
\caption{Simulated background due to the decay of the two daughters $^{214}$Bi($\beta^-$) and 
$^{214}$Pb($\beta^-$) of the $^{226}$Ra decay chain for 10$^5$ decays of each isotope. The events below
1~MeV are mainly due to the decay of $^{214}$Pb with a Q-value of 1.03~MeV, while all the events above 1~MeV 
correspond to decays of $^{214}$Bi with a Q-value of 3.27~MeV.}
\label{fig_214}
\end{figure}

During the runs carried out at DANCE a significant background component between 0.5 and 3.5~MeV was 
discovered,
which could not be discriminated via particle identification (see Fig.~\ref{fig_bg}). The observed count rate
per crystal is about the same as the $\alpha$-activity. As a result of the 
$^{226}$Ra inside the BaF$_2$, the crystals contain not only $\alpha$, but also $\beta^-$ activities. The 
most important contributions are coming from $^{214}$Bi($\beta^-$) ($Q_\beta=3.3$~MeV) and 
$^{214}$Pb($\beta^-$)
($Q_\beta=1.0$~MeV). All other decays either release energies below 100~keV or are not yet in equilibrium, 
since  
$^{210}$Pb with a half life of 22.3~years acts as a bottle neck in the decay chain. In order to confirm these
decays as the cause of the observed background, GEANT simulations have been carried out. In a first step, 
decay cascades according to the (in the case of $^{214}$Bi($\beta^-$) rather complex) decay scheme 
\cite{BrF86,Fir96} 
have been created using a Monte Carlo method. The energy distribution of the emitted electrons was calculated
by means of the Fermi theory of beta decay. In a second step, these cascades were started inside the BaF$_2$
scintillators and the response of the array was simulated. Fig.~\ref{fig_214} shows the result 
for 10$^5$ decays of $^{214}$Bi and $^{214}$Pb. The agreement between the experimental data 
(Fig.~\ref{fig_bg})
and the corresponding spectrum in the middle section of Fig.~\ref{fig_214} is remarkable. The bottom part of 
Fig.~\ref{fig_214} shows the same data with a cluster multiplicity cut of greater than 1 applied. A cluster is 
defined as a set of neighboring fired crystals, completely surrounded by crystals that
register no hits. The simulations suggest that this background can be significantly discriminated using such a
cluster analysis. While the total detection efficiency for a $^{214}$Bi or $^{214}$Pb event is above 99\%, 
the probability for events creating more than 1 cluster is 16\% or 3\%, respectively. For typical neutron 
capture events however this would be more than 80\% \cite{RHK01}.  

\subsection{Time-dependent background}
\label{time-dependent}
Time-dependent background is correlated with the time structure of the neutron beam, but does not scale
with the size of the sample. At pulsed neutron sources the neutrons are usually preceded by other particles, 
e.g. the so-called $\gamma$-pulse ($\gamma$-flash), originating from the interaction of the primary 
beam with the neutron production target. Because FP-14 is designed to view an upper tier water moderator 
and not the tungsten cylinder directly, the DANCE detector is shielded against primary particles from 
the production target. Additionally the apertures used for neutron collimation restrict the geometrical 
size of the flow for most of the background
particles coming along the flight path. Therefore, without any material in the neutron beam almost no 
prompt $\gamma$-flash would be observed 
with the DANCE array. 

\begin{figure}
\begin{center}
\includegraphics[width=.9\textwidth]{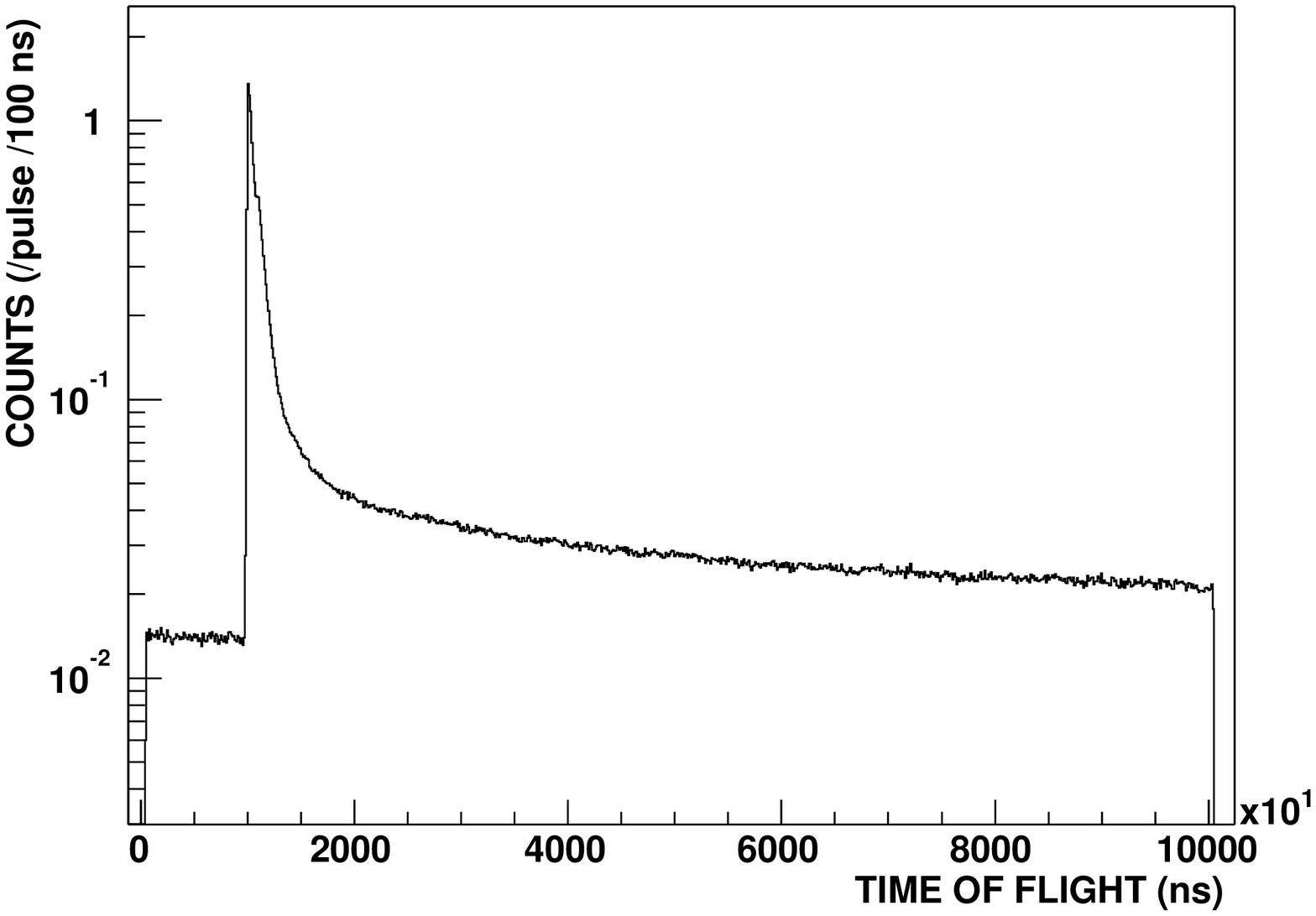}
\end{center}
\caption{
Beam-related background measured at DANCE. The first 10~$\mu$s are preceding the actual 
proton pulse and characterize the time-independent background. No multiplicity or energy cut has been applied.
The right end of the TOF spectrum corresponds to a neutron energy of 260~eV. With the estimate given in 
Sect. \ref{count_estimate}, the count rate for neutron capture would be $10^{-4}$ in this plot. 
In this case of no energy or multiplicity cuts, the background rate is 2 orders
of magnitude above the neutron capture rate.
}
\label{fig_tof_bkg}
\end{figure}

Nevertheless, due to practical reasons a set of windows is installed on FP-14 in order to
divide the beam pipe volume into smaller segments. Furthermore a neutron monitor is installed downstream of 
the 
DANCE array. Neutrons as well as $\gamma$-rays (the most important second
component in the beam) interact with these windows and cause time-dependent background in the BaF$_2$ 
detector.
Fig.~\ref{fig_tof_bkg} shows the number of events per pulse as a function of time of flight without any 
sample in the beam.

\begin{figure}
\begin{center}
\includegraphics[width=.9\textwidth]{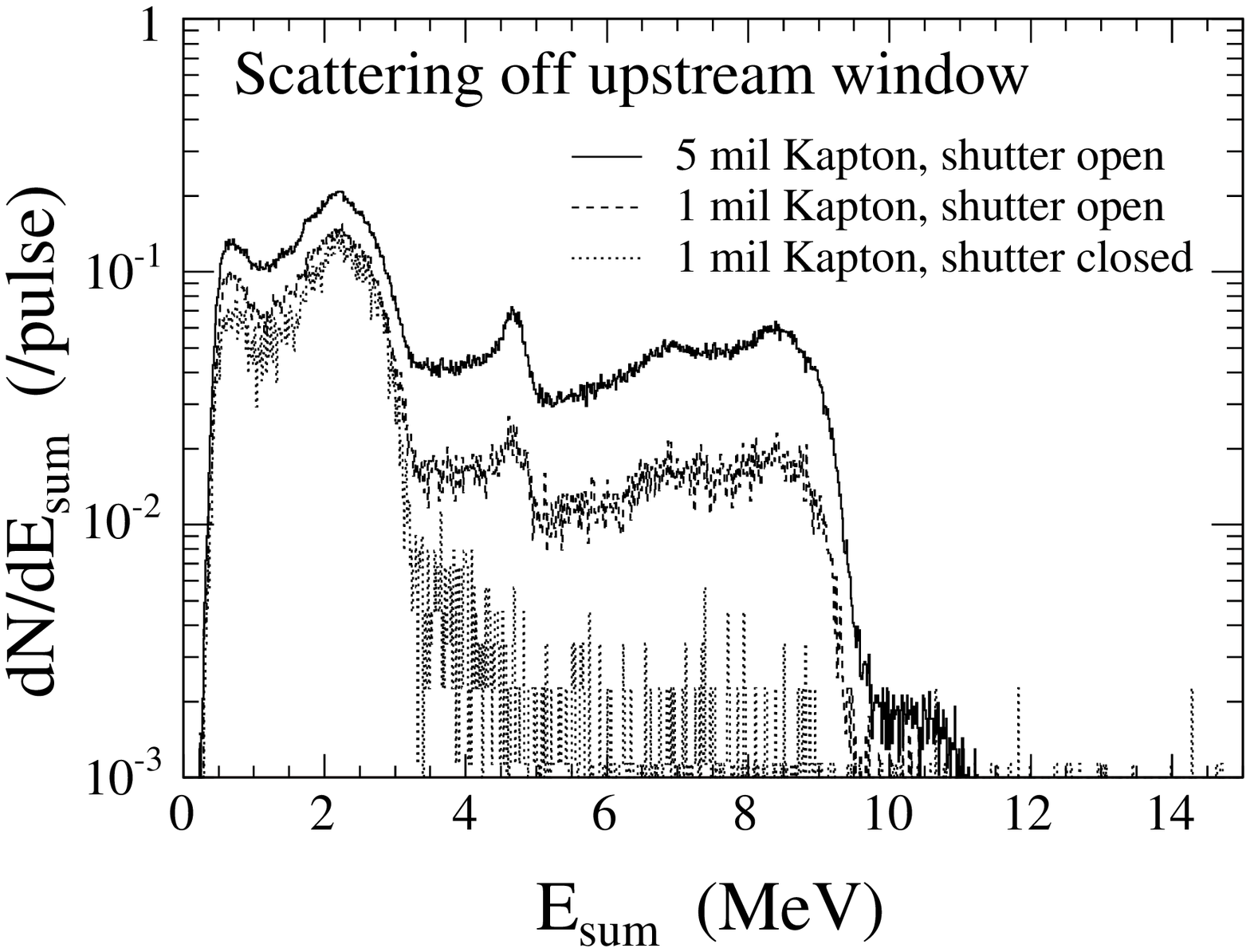}
\end{center}
\caption{Example for background reduction achieved by optimizing beam pipe windows. Histogram 
binning is 15~keV/channel. One pulse corresponds to 1.6~ms looking time. The rate estimate given in 
Sect. \ref{count_estimate} is $2.4\cdot10^{-2}$ for this plot, which is above the background rate
for energies of 3~MeV or higher. This shows that the 
discussed background reduction was sufficient. }
\label{bkgd_windows}
\end{figure}

Based on Monte Carlo simulations and the observed rather high beam induced background all the previously used, 
rather thick, Al-windows inside the experimental area have been replaced by KAPTON foils of 25~$\mu$m (1 mil) 
thickness and the neutron monitor has been moved further downstream. 
Both changes significantly reduced the time-dependent background (see Fig.~\ref{bkgd_windows}).

\subsection{Sample-related background}
\label{}
For the purpose of this article only background created by the interaction of particles in the beam 
with the sample are under investigation. The possible radioactivity of the sample is not discussed here.

\subsubsection{Neutron scattering reactions}
\label{}
\begin{figure}
\begin{center}
\includegraphics[width=.9\textwidth]{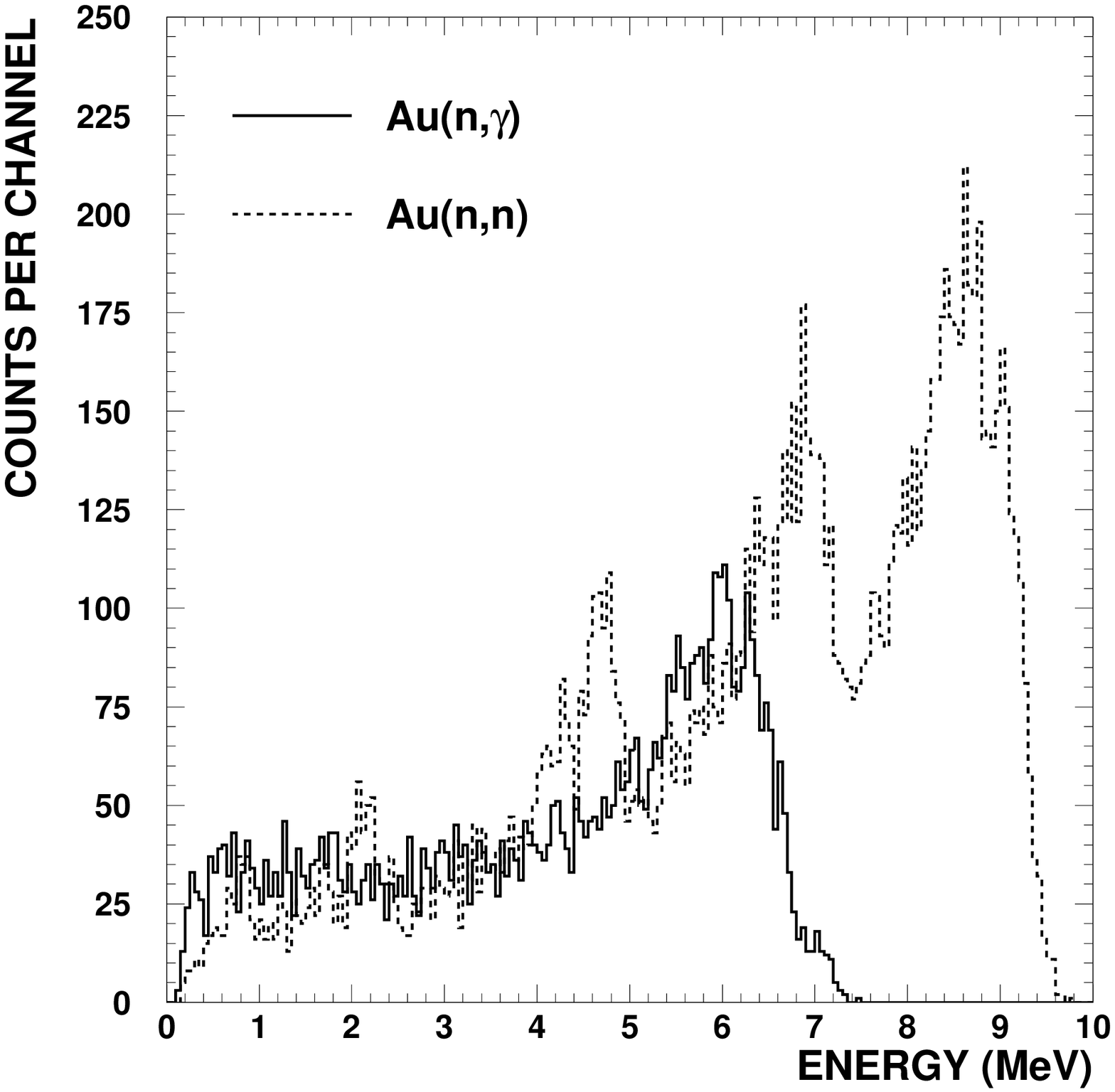}
\end{center}
\caption{Simulated total energy spectra for the DANCE array with 141~crystals and high single
detector threshold without the $^6$LiH absorber. $10^7$ neutrons between 10~keV and 100~keV
have been emitted towards a
0.2~mm thick gold sample. The solid line corresponds to events due to captures on a gold sample, 
the dashed curve to events due to neutrons scattered at the gold sample and captured on the isotopes
of Ba and F. Histogram binning is 50~keV/channel.}
\label{fig_au_ng}
\end{figure}

The scattering cross section for neutrons with keV energies is typically 1-2 orders of magnitude
larger than the capture cross section. Therefore, this unavoidable background component might dominate the 
experiment.
Even though the concept of a 100\% efficiency detector in principle allows discrimination between neutron 
captures 
on different isotopes, further reduction of background due to scattered neutrons is necessary.
The significance of this problem has been explored by means of Monte-Carlo simulations of neutrons on a gold 
sample
using GEANT 3.21 \cite{RBB03a}, as shown in Fig.~\ref{fig_au_ng}.

\begin{figure}
\begin{center}
\includegraphics[width=.9\textwidth]{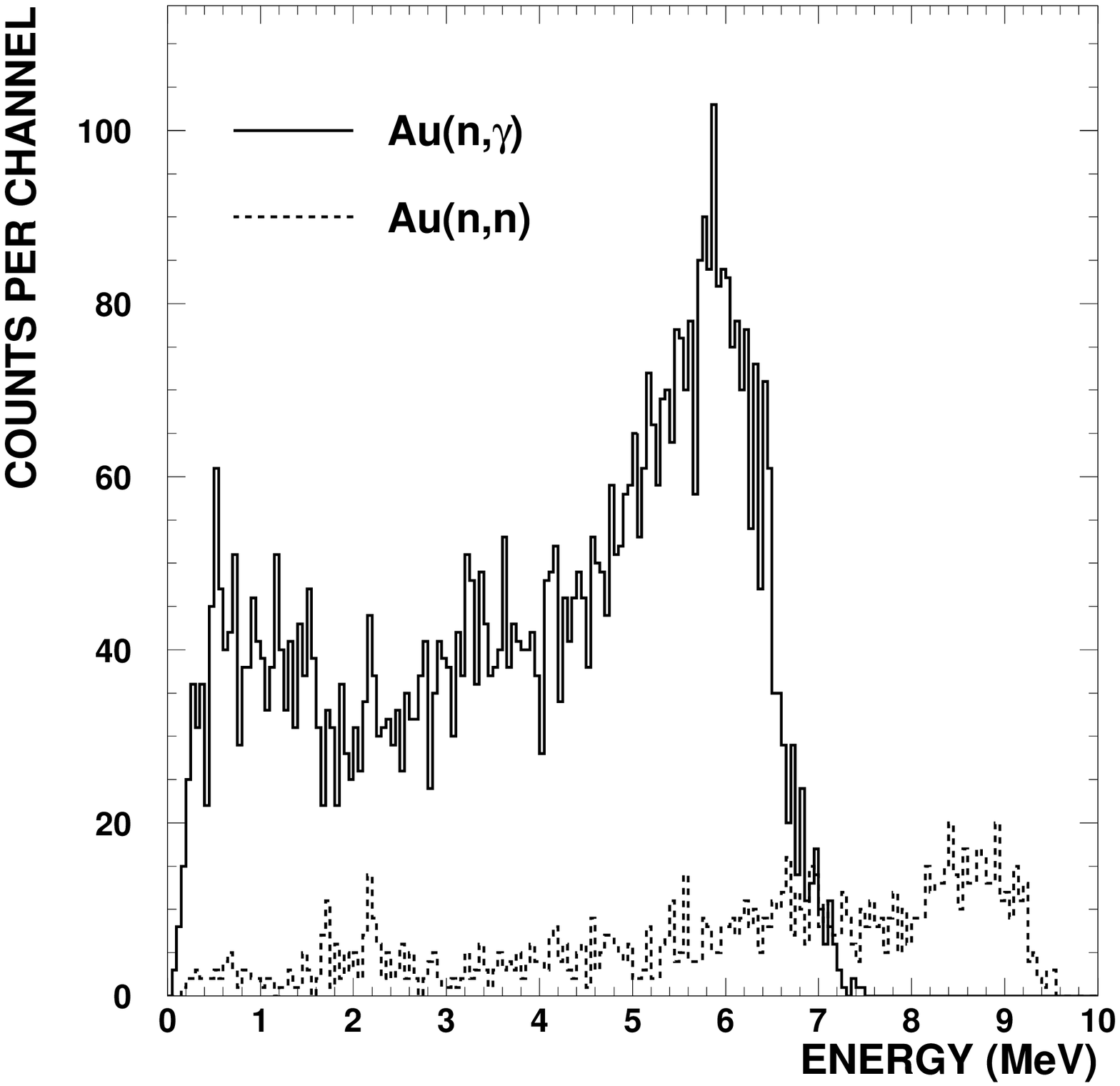}
\end{center}
\caption{Simulated total energy spectra for the DANCE array with 141~crystals and high 
single detector threshold with the $^6$LiH absorber. $10^7$ neutrons between 10~keV and 100~keV
have been started towards a
0.2~mm thick gold sample. The solid line corresponds to events due to captures on a gold sample, 
the dashed curve to events due to neutrons scattered at the gold sample and captured on the isotopes
of Ba and F.  Histogram binning is 50~keV/channel.}
\label{fig_au_ng_lih}
\end{figure}

\begin{figure}
\begin{center}
\includegraphics[width=.9\textwidth]{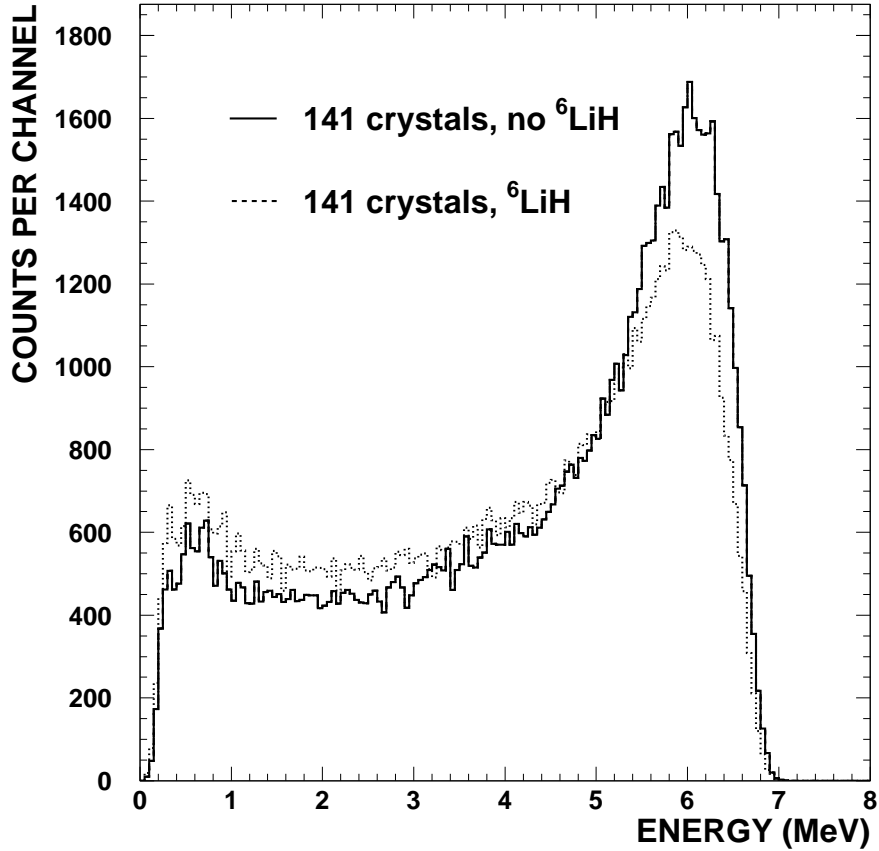}
\end{center}
\caption{Simulated total energy spectra for the DANCE array with 141 crystals and high single detector
thresholds for gamma radiation from neutron capture on gold. $10^5$~cascades were simulated. The histogram 
binning is 50~keV/channel. For details about the capture cascades used for these simulations see Ref. 
\cite{UhK93}.}
\label{fig_au_ng_lih_peak}
\end{figure}

\begin{figure}
\begin{center}
\includegraphics[width=.9\textwidth]{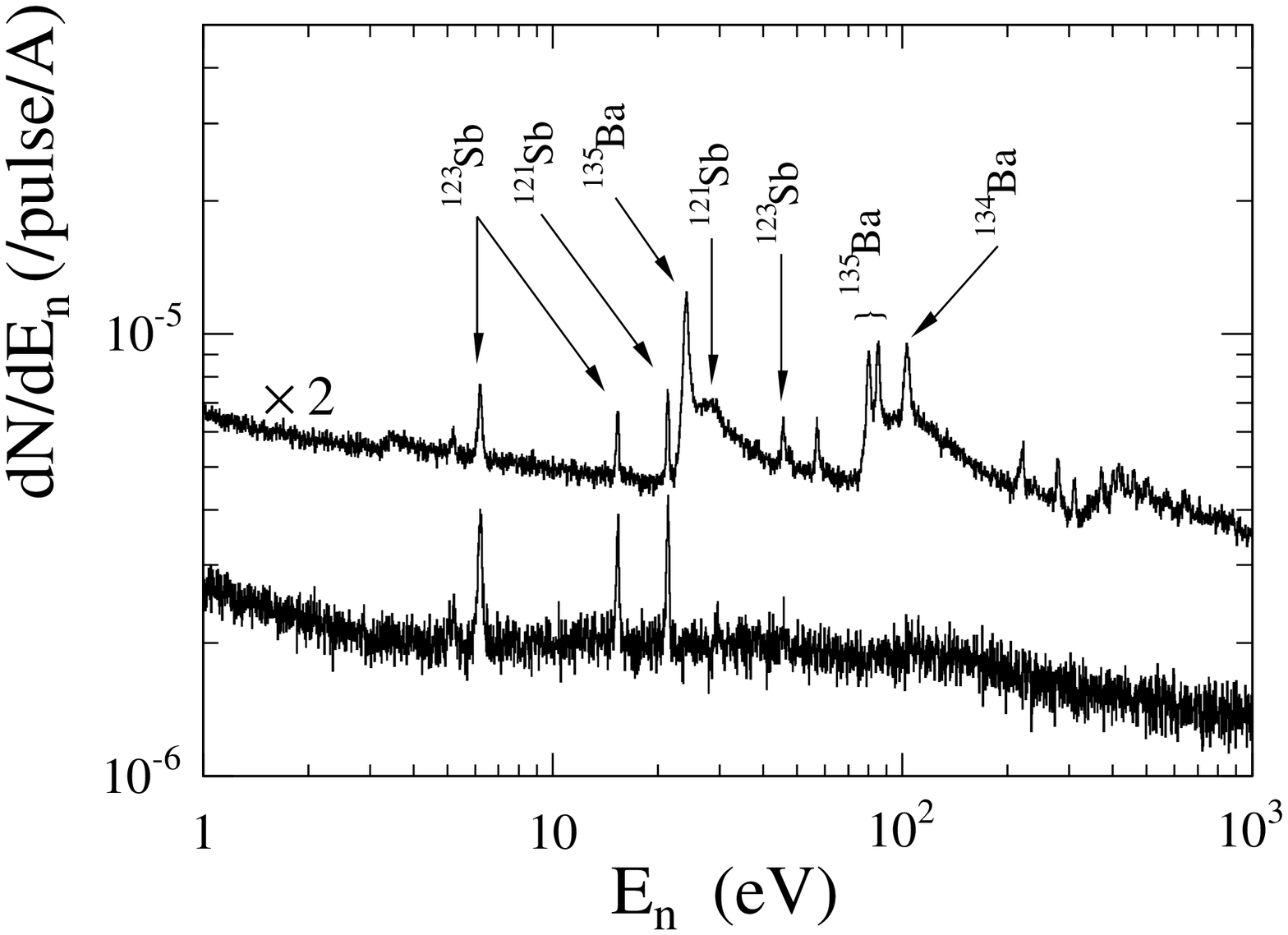}
\end{center}
\caption{Events detected with DANCE as a function of incident neutron energy. A 43.5~mg/cm$^2$ lead sample
was in the beam and 141 crystals were connected. Top curve: Without $^6$LiH absorber
Bottom curve: With $^6$LiH absorber. The spectra are normalized to proton beam current (in ampere) and pulse. 
The binning of the x-axis is logarithmic and contains 1000 channels per energy decade. Only a single detector 
energy threshold of about 400~keV was applied.}
\label{fig_pb_lih}
\end{figure}

The DANCE collaboration decided to use a passive method of reducing the number of neutrons scattered into the
surrounding materials. A $^6$LiH shell with an inner radius of 10.5 cm and an outer 
radius of 16.5~cm was installed inside the DANCE ball. This shell efficiently moderates and absorbs neutrons 
scattered at the sample without producing $\gamma$-rays (see Fig.~\ref{fig_au_ng_lih} and 
Table~\ref{tab_scatter_capture}). 
Furthermore, the material has been chosen to be as transparent as possible for gamma radiation originating 
from neutron 
captures on the sample. Nonetheless, the simulations predicted a reduction of the sum peak efficiency 
(Fig.~\ref{fig_au_ng_lih_peak}). The experimental effect of the $^6$LiH shell can be seen in 
Fig.~\ref{fig_pb_lih}, where two runs with a lead sample are compared. Natural lead predominantly scatters 
neutrons.
The peaks due to resonant capture of scattered neutrons on the surrounding barium are at least a factor of 20 
reduced after installing the $^6$LiH shell, while the peaks due to resonant captures on the Sb impurities
inside the 
lead sample remain at the same level.     

\begin{table}
 \caption{Ratio of events from scattered neutrons and capture events on the sample for different setups and 
neutron energy
 intervals. The simulations contained 141 BaF$_2$ crystals and a high detector threshold. The last
column is for events with total deposited energy above 1 MeV only, while all other ratios correspond to the 
total
number of detected events. The last line shows the ratio between the two first lines.} 
   \label{tab_scatter_capture}
   \begin{tabular}{lcccc}
	 Setup & 0.1 .. 1 keV & 1 .. 10 keV & 10 .. 100 keV & 0.1 .. 1 MeV \\
    \hline
    no $^6$LiH 	 	  	  	  	 & 0.47    & 1.2 	  & 2.2    & 2.7 		 \\
	$^6$LiH	 					 & 0.0035  & 0.025	  & 0.20   & 0.66		 \\
	Ratio 						 & 134	   & 48		  & 11	   & 4.1		 \\
   \end{tabular}
\end{table}

\subsubsection{$\gamma$-induced reactions}
\label{gamma_induced}
Experiments and FLUKA simulations at n-TOF as well as GEANT and MCNP simulations for target 1 at 
WNR show that about 1 $\gamma$-ray per 10 neutrons will be present at the sample position in the keV-neutron 
TOF region. The detection efficiency of the DANCE array for photons is about 90\%, while it is 
only 10\% for scattered neutrons (without the $^6$LiH absorber shell). Thus 
the potential ratio of events caused by scattered
neutrons to events by scattered photons is about equal, before weighting with the
respective cross sections due to interaction with the isotope at the sample
position. 

The $\gamma$-ray spectrum depends strongly on the reaction producing the photons as well as
on the specifics of the setup.
The $\gamma$-spectrum at late times (long time of flight) consists mainly of two components:\\
- prompt photons due to capture of moderated neutrons \\
- photons due to the decay of (neutron induced) reaction products

\begin{figure}
\begin{center}
\includegraphics[width=.9\textwidth]{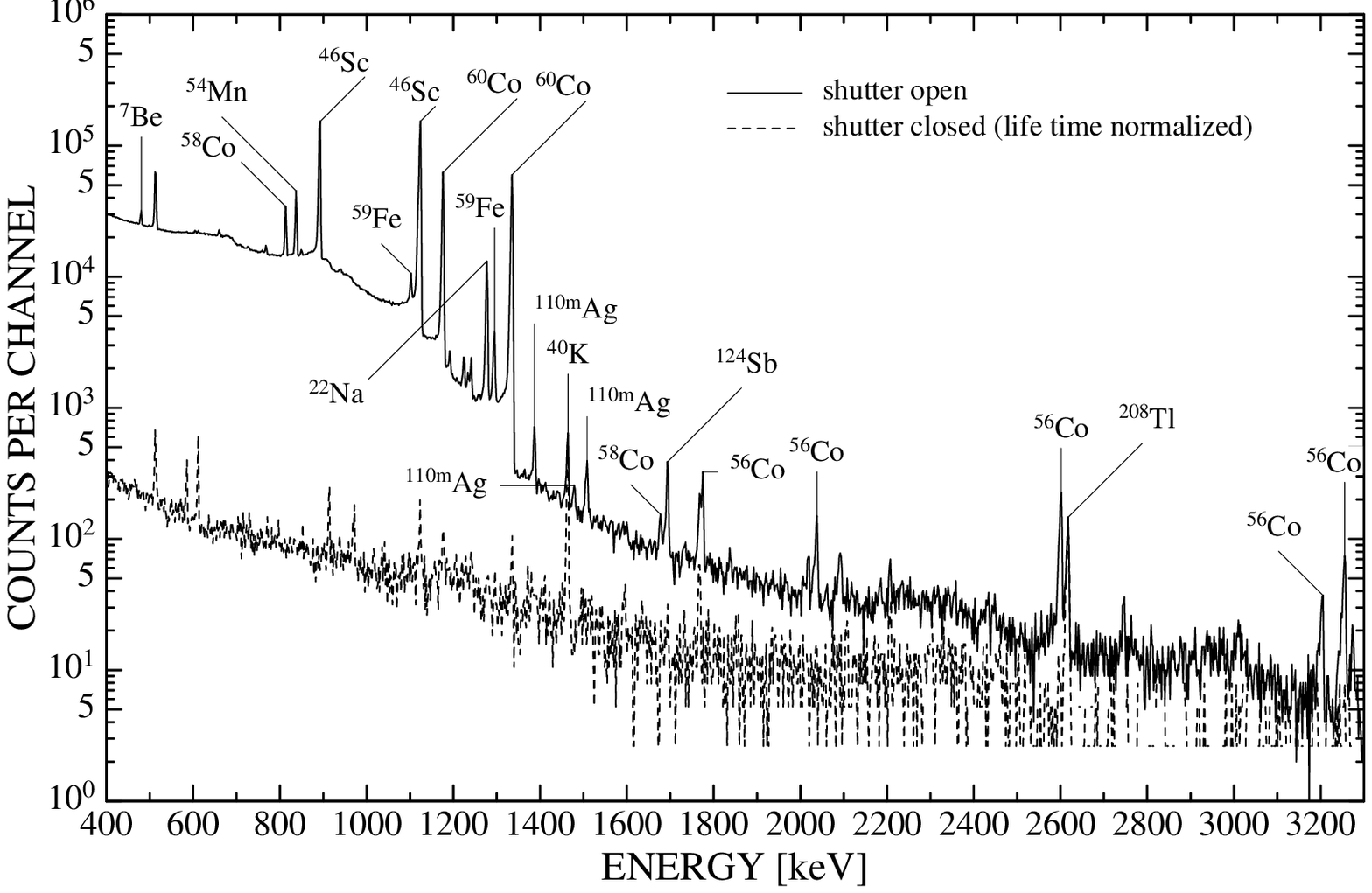}
\end{center}
\caption{Gamma-ray spectrum measured with a germanium detector 
at flight path 14 at the Manual Lujan Center at LANL 3~months 
after the last neutron beam. Both runs are normalized to the same life time.}
\label{fig_fp14_bg}
\end{figure}

The component due to the decay of reaction products was measured at FP14 with a germanium detector 
three months after the shutdown 
of the accelerator (see Fig.~\ref{fig_fp14_bg}). All the observed lines correspond to activation due to 
neutron induced reactions on aluminum alloy and stainless steel. Both materials are present in windows 
upstream of the shutter, and the water moderator container is made from aluminum alloy.

The capture $\gamma$-ray component has not been measured directly yet. Since both neutron as well as gamma 
flux during accelerator operations is very high, it is almost impossible to place any kind 
of detector directly in the beam. Indirect methods need to be applied. Depending on the atomic number
of a sample in the beam, the total interaction cross section varies significantly as a function of the photon 
energy. 
The DANCE array is a powerful tool to detect the $\gamma$-ray component of such interactions, since almost all 
of the secondary $\gamma$-rays will be absorbed in the BaF$_2$ crystals.

\begin{figure}
\begin{center}
\includegraphics[width=.9\textwidth]{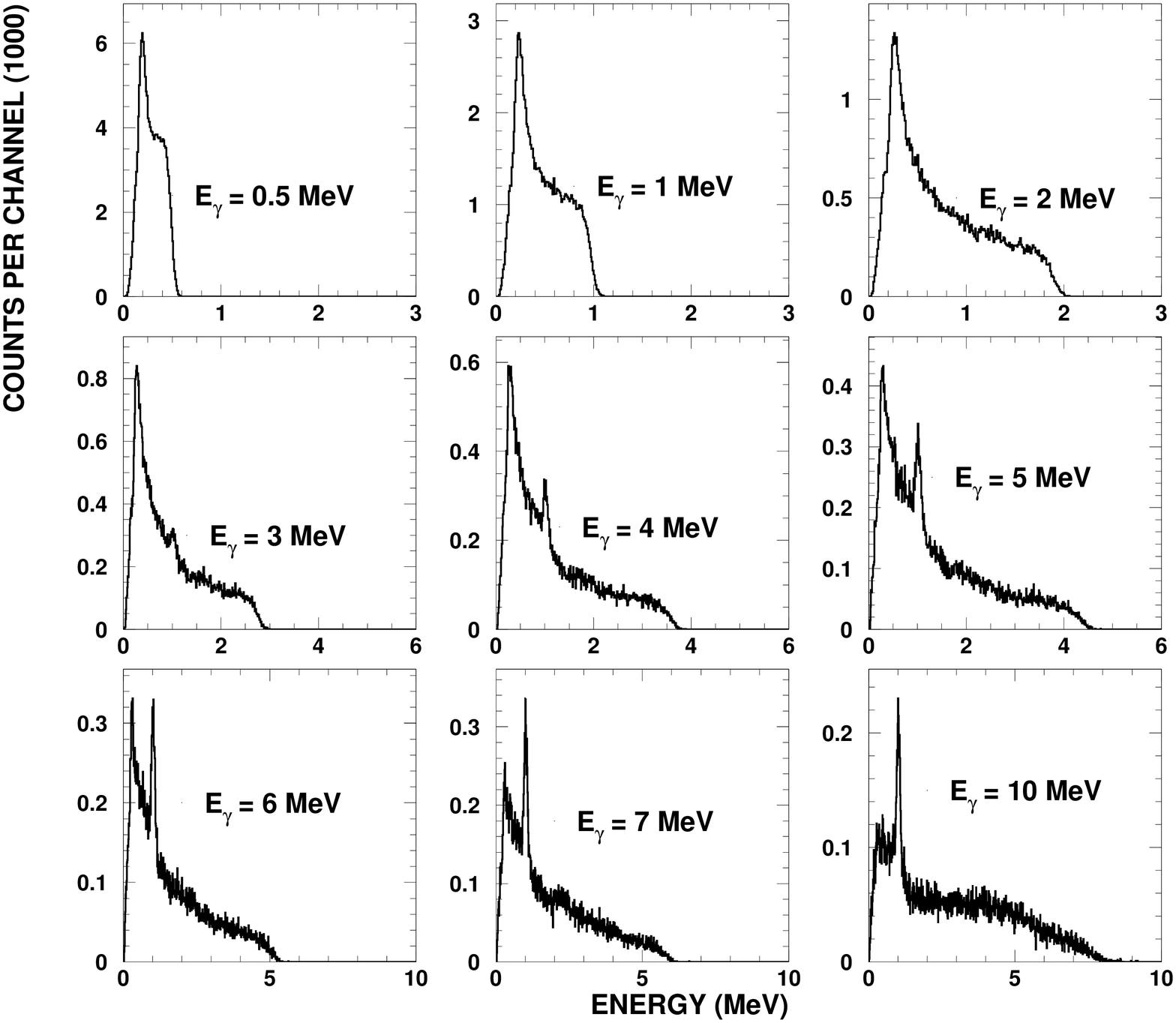}
\end{center}
\caption{Simulated total energy deposition in the BaF$_2$-array for $10^6~\gamma$-rays of different energies 
interacting with a carbon sample (10~mm thickness) in the center of the detector. The histogram binning is 
10~keV/channel. For more detailed information see Ref. \cite{RBB03b}.}
\label{fig_in_beam_c}
\end{figure}

\begin{figure}
\begin{center}
\includegraphics[width=.9\textwidth]{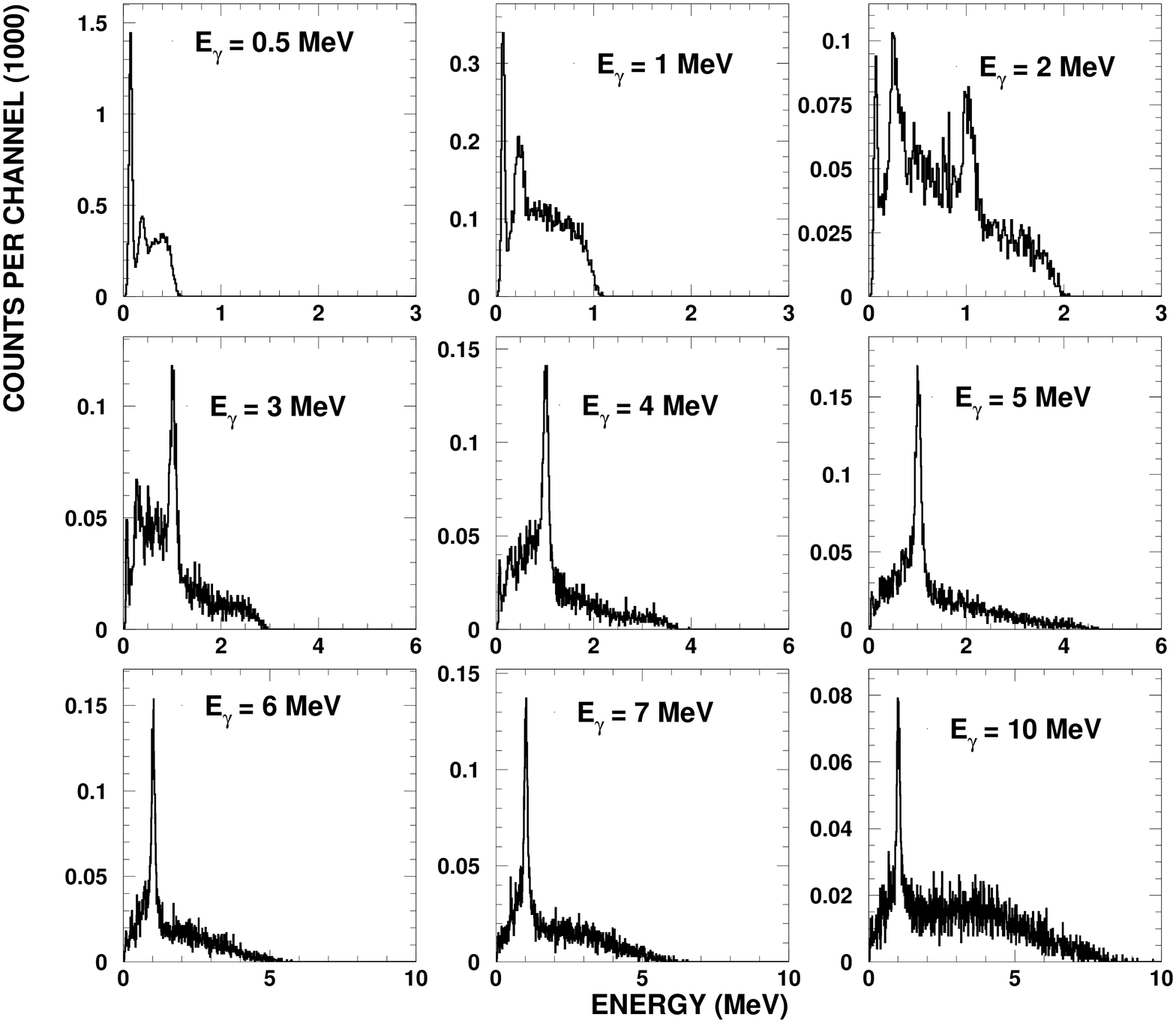}
\end{center}
\caption{Simulated total energy deposition in the BaF$_2$-array for $10^6~\gamma$-rays of different energies 
interacting with a lead sample (0.2~mm thickness) in the center of the detector. The histogram binning is 
10~keV/channel. 
For more detailed information see Ref. \cite{RBB03b}.}
\label{fig_in_beam_pb}
\end{figure}

Carbon is often used for determining the neutron scattering background, since the neutron
capture cross section is orders of magnitude smaller than the scattering cross section. Fig.
\ref{fig_in_beam_c} shows a simulated response of the 4$\pi$ detector to photons on a carbon sample, 
typical for samples with low atomic number. 
The most important interaction mechanisms in this case are photo effect and Compton scattering. Only 
high-energy photons interact primarily due to pair production. Therefore, the
1.02~MeV peak due to pair production can only be observed for primary photons above 4~MeV.
As mentioned previously, another important neutron scattering sample is lead. In contrast to the case of 
carbon, 
the spectra gained for a lead sample reflect the typical for photon interaction with heavy nuclei 
(Fig.~\ref{fig_in_beam_pb}). Pair production becomes 
the most important interaction, resulting in a peak at 1.02~MeV. For more detailed information on those 
simulations see also Ref. \cite{RBB03b}.

\begin{figure}
\begin{center}
\includegraphics[width=.9\textwidth]{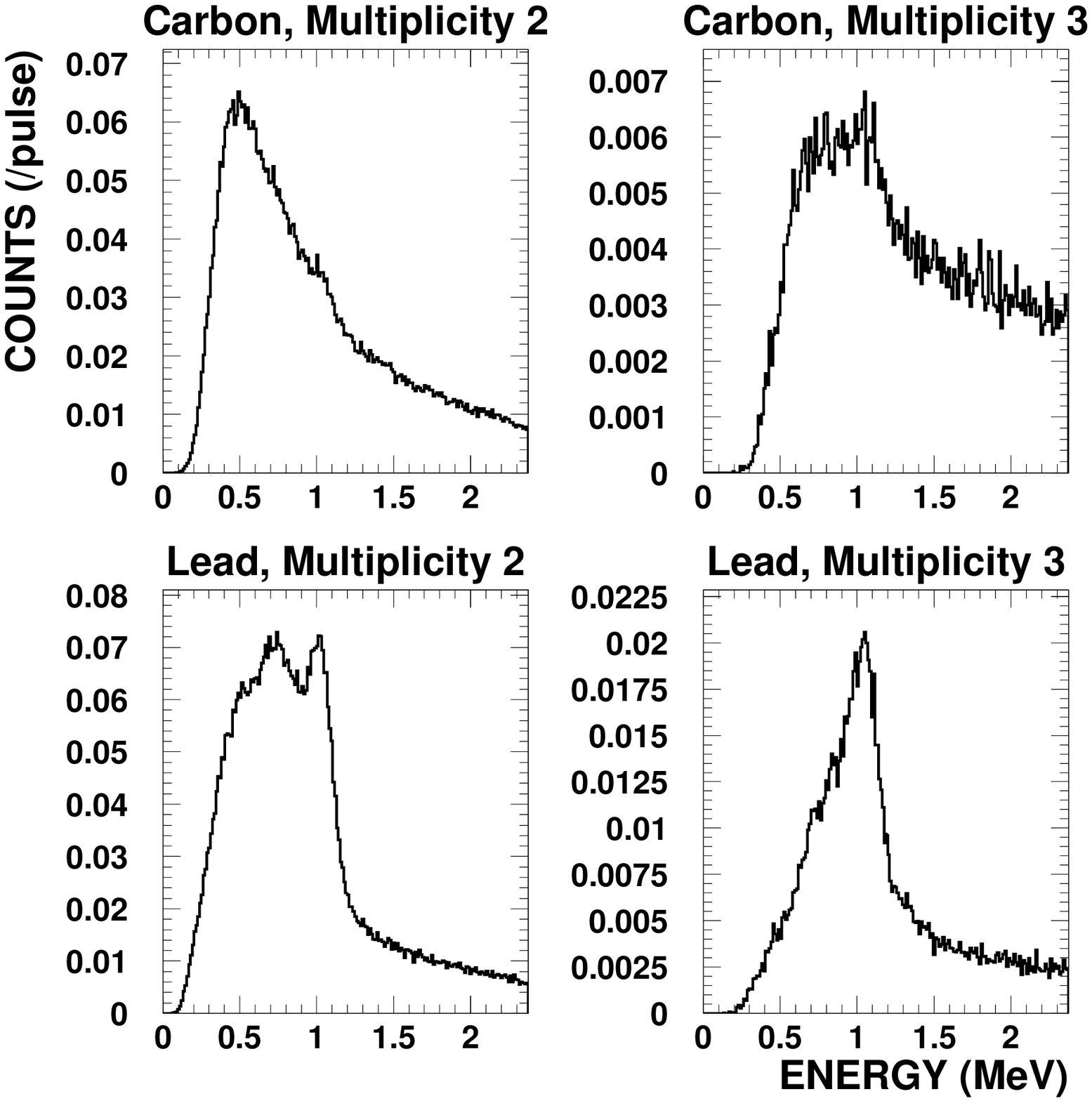}
\end{center}
\caption{Total energy spectra for a C and a Pb sample taken with the DANCE array. All measurements have 
been carried out for 0.5~h with a 1.3~cm
polyethylene filter in the beam 10~m upstream of the sample. The top panels show the results for a
44.4~mg/cm$^2$
carbon sample and the bottom for a 43.5~mg/cm$^2$ lead sample. The left and right pictures correspond to
crystal 
multiplicity 2 and 3, respectively. The TOF interval is the same as in Fig.~\ref{fig_tof_bkg}. The histogram
binning is 12~keV/channel. One pulse corresponds to 100~$\mu$s looking time.}
\label{fig_in_beam_c_pb_zoom}
\end{figure}

\begin{figure}
\begin{center}
\includegraphics[width=.9\textwidth]{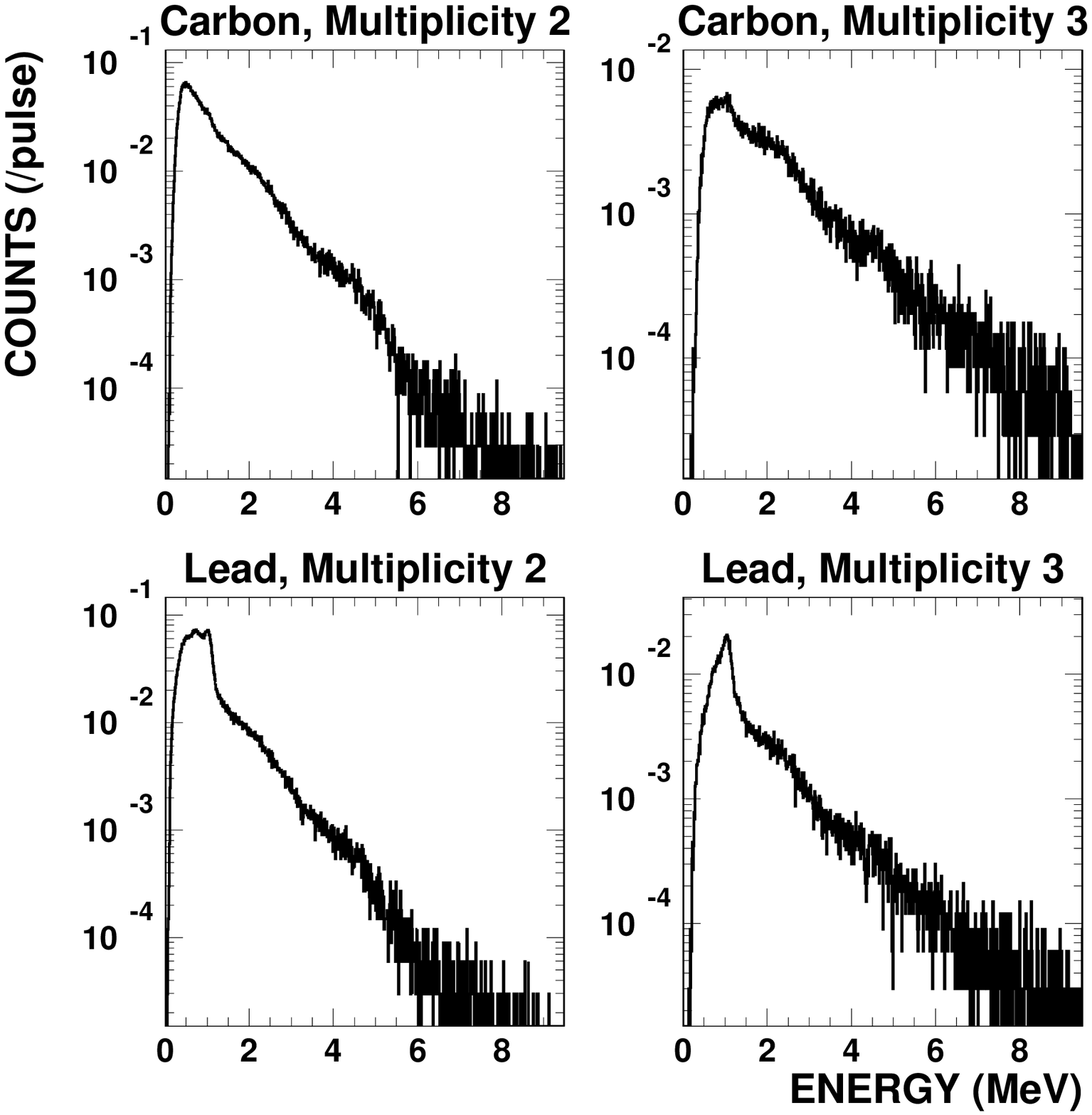}
\end{center}
\caption{The same data as in Fig.~\ref{fig_in_beam_c_pb_zoom}, but with different energy scale and
logarithmic y-axis.}
\label{fig_in_beam_c_pb_log}
\end{figure}

Figs.~\ref{fig_in_beam_c_pb_zoom} and~\ref{fig_in_beam_c_pb_log} show measured spectra, which
underline the significance of this background component. For these runs a carbon and a lead sample 
have been placed in the beam and the spectra have been taken with the DANCE array. The TOF interval is
the same as in Fig.~\ref{fig_tof_bkg}, corresponding to neutron energies of 0.26~keV or higher. 
A 1.3 cm thick absorber of polyethylene
was placed approximately in the middle between the neutron production target and the sample, which reduced the
total neutron flux at the sample below 100~keV by at least a factor of 10, while the $\gamma$-ray flux is
reduced less than 5\%. Therefore most of the events shown in the figures are related to photon interactions with 
the sample. Especially in the case of lead, the peak at 1.02~MeV can be clearly recognized. The absence of 
this peak 
in the case of carbon suggests a $\gamma$-ray spectrum with energies
predominantly between the pair production threshold (since lead shows the 1.02~MeV peak) and 
3~MeV (otherwise carbon should show the 1.02~MeV peak too). A possible 
production mechanism could be H(n,$\gamma$) in the moderator, resulting in a $\gamma$-ray with an energy of 
2.2~MeV.
More detailed investigations will be carried out with the completed array.

\section{Conclusions}
\label{}
Modern neutron capture experiments require highly efficient detectors to enable
the use of small samples, hence the investigation of rare or radioactive isotopes. 
In order to achieve a sufficient
signal to background ratio, background suppression has to be emphasized and
residual background components have to be well understood. During its first beam
period the DANCE experiment concentrated on identifying, understanding and 
suppressing the applicable background components. This work was 
assisted by intensive Monte Carlo simulations. 

As a result of these efforts the intrinsic $\alpha$-background can now be well 
separated from $\gamma$-events.
In later stages these suppressed events will be used to automatically stabilize 
the high voltage of the PMT, which should in turn stabilize the energy calibration. 
Time-dependent background components induced
by structural material in the beam line and shielding have been reduced by optimized choices
of materials and dimensions. As expected from simulations, the installation of a $^6$LiH absorber
between sample and detector sphere reduces the background from scattered neutrons significantly. 
Although reducing the peak efficiency
the use of this absorber shell clearly improves the signal to background ratio.

Gamma induced background will be further investigated, since it is shown to be an
important component. Starting in Summer 2003 neutron capture measurements on a number
of stable as well as unstable isotopes have been successfully carried out the DANCE. 
The analysis of these experiments is in progress.

\ack{
We would like to thank L.F.~Hunt, E.P.~Chamberlin, D.R.~Harkleroad, N.F.~Archuleta and D.M.~Lujan 
for their invaluable help during the development and construction phase of DANCE. Special
thanks also to M.~Heil and F.~K\"appeler from FZK for countless clarifying discussions.
This work has benefited from the use of the Los Alamos Neutron Science Center at the Los Alamos 
National Laboratory. This facility is funded by the US Department of Energy and operated 
by the University of California under Contract W-7405-ENG-36. The Colorado School of Mines group is funded via 
DOE grant: DE-FG02-93ER40789.
}  

\newcommand{\noopsort}[1]{} \newcommand{\printfirst}[2]{#1}
  \newcommand{\singleletter}[1]{#1} \newcommand{\swithchargs}[2]{#2#1}

\end{document}